\documentstyle[amssymb,12pt]{article}

\input{tcilatex}

\begin{document}

\author{Anthony Rizzi \\
{\small {\it Institute for Advanced Physics}}}
\title{The Meaning of Bell's Theorem}
\date{}
\maketitle

\begin{abstract}
The import of Bell's Theorem is elucidated. The theorem's proof is
illustrated both heuristically and in mathematical detail in a pedagogical
fashion. In the same fashion, it is shown that the proof is correct
mathematically, but it doesn't require, as is usually thought, one to
abandon locality or realism.

\end{abstract}

\section{Introduction}


\qquad Bell's 1964 paper\cite{BellEPR64} addressing the Einstein, Podolsky,
Rosen's (EPR) paradox\cite{EPR1935} peaked the debate between supporters of
locality and/or realism on the one hand and the so-called orthodox, or
Copenhagen,\footnote{%
At its most generic, no one objects to the Copenhagen interpretation, for at
that level it only expresses the complementarity of wave-particle aspects
found in measurements. That is, at that level, it simply points to the fact
that one type of measurement manifests wave aspects and another particle
aspects. However, usually the Copenhagen interpretation carries, at least
implicitly, much more philosophical baggage than this. It has, for example,
been taken to include the negation of the principle of causality.}
interpretation of quantum mechanics on the other.\footnote{%
In this paper, I do not discuss the Everett's so-called many worlds
interpretation of Quantum mechanics.} Here {\it locality} is the condition
that one physical thing cannot influence another except by propagation of
the effects in a finite time. {\it Realism} is the idea that the physical
world, though we can affect it, exists fully independent of us. Both seem
benign, indeed trivially true, to those not acquainted with quantum
mechanics. However, for many familiar with quantum mechanics, the debate was
capped in favor of Copenhagen when, in the early 80's, Alain Aspect did a
series of experiments that confirmed that quantum mechanics accurately
described the physical situation of the EPR paradox.\cite{AspectThesis83} 
\cite{Aspect82}

At the heart of the controversy stirred by Bell's theorem lies a continuing
debate about the interpretation of the mathematical formalism of quantum
mechanics. In the earliest days of the quantum theory, grappling with the
interpretational problems plaguing the theory led to the traditional
Copenhagen interpretation - a synthesis of Bohr's and Heisenberg's views
(although neither of them used this term to refer to their joint views).\cite
{Bohrvol1}\cite{Bohrvol2}\cite{Bohrvol3}\cite{Bohrvol4} Even then, there
were important differences between Bohr's view and Heisenberg's view that
were never completely resolved. Heisenberg's view was more subjectivistic
than Bohr's; he initially took the uncertainty relations that he derived to
imply that certain objects do not exist in nature unless and until we
observe them. This is not surprising considering the zeitgeist of the time
in Germany. Contrary to Heisenberg, Bohr rejected the view that the
experimental outcome is due to the observer. Instead, he chose to view the
uncertainty principle in the larger framework of what he called
complementarity. Bohr's complementarity was not particularly positivistic or
subjectivistic; to a certain extent this is because he was quite vague about
its descriptive content. Philosophically, he was also influenced by
neo-Kantianism.\cite{stanfordEncycl02} As evidenced by Bohr's conviction
that the atom is real, he had an inkling that the natural world exists
independently of experiments; however, precisely what sort of inkling he had
is, at best, unclear. The subjectivist interpretation of the quantum theory
set off one of Einstein's quotable remarks. He asked his friend A. Pais, if
he ``really believed that the moon exists only when [you] look at it?'' \cite
{MerminNotreD}

Bell's theorem is controversial because it purports to show that physical
reality must be non-local; furthermore some use it to bolster the
subjectivist interpretation of quantum mechanics which claims a thing isn't
there until you measure it. For Bell, his renowned theorem shows that even
if one takes quantum mechanics to be incomplete and tries to expand it by
use of ``hidden variables'' one must have, what Einstein called ``spooky''
action at a distance. We will see, by taking seriously the implications of
the fact that all non-commuting observables are not simultaneously
measurable, that one need not sacrifice locality.

\section{Bell's Theorem and the EPR Paradox}

For completeness and maximal usefulness for instruction, even at an
introductory level, this section has three parts. First, to give the
complete ground work for Bell's proof, the predictions of quantum mechanics
are summarized for correlated spin 1/2 particles. The second part gives a
concrete description of the EPR\ experiment and Bell's proof; it makes clear
why the correlation predicted in the first part is inconsistent with Bell's
inequality and hence apparently with locality as well. Again, such a
non-locality means that a thing here instantaneously influences something on
the other side of the universe. The final part gives the mathematical proof,
relating it to the description and heuristic arguments of the second part.

\subsection{The Quantum Mechanics of a pair of electrons in the Singlet State%
}

In quantum mechanics, spin is a 3-vector quantity ${\bf S}$.\footnote{%
In the following section we shall be using finite-dimensional Hilbert spaces
to describe spin angular momentum of quantum mechanical systems.} We have $%
S^{2}=S_{x}^{2}+S_{y}^{2}+S_{z}^{2}$ as well as the commutation relations $%
[S_{x},S_{y}]\equiv S_{x}S_{y}-S_{y}S_{x}=iS_{z}$ (and other permutations of 
$\{x,y,z\}$) where we choose units so that $\hbar =1$.

Furthermore, we define the usual \ ``ladder'' operators $S_{+}$ and $S_{-}$
such that 
\begin{eqnarray}
S_{x} &=&1/2(S_{+}+S_{-})\qquad \\
S_{y} &=&-i/2(S_{+}-S_{-}).  \nonumber
\end{eqnarray}

Then, the following result holds: $S_{\pm}\mid S_{z}=m > = (s(s + 1) - m(m
\pm 1)) \mid S_{z} = m \pm 1>$.

We now define a new operator ${\bf \sigma }$ = $2{\bf S}$ and note that the
eigenvalues of $\sigma _{x},\sigma _{y},\sigma _{z}$ are $\pm 1$. It is
convenient to introduce more notation at this point:

\begin{eqnarray}
&\mid &\alpha >\,=\,\mid \sigma _{z}\,=+1>\qquad \\
&\mid &\beta >\,=\,\mid \sigma _{z}=-1>  \nonumber
\end{eqnarray}

Projecting the $\sigma$-components onto the $\{ \mid\alpha>,\mid\beta> \}$
basis yields the representation

\begin{equation}
\sigma _{x}=\left( 
\begin{array}{cc}
0 & 1 \\ 
1 & 0
\end{array}
\right) ,\sigma _{y}=\left( 
\begin{array}{cc}
0 & -i \\ 
i & 0
\end{array}
\right) ,\sigma _{z}=\left( 
\begin{array}{cc}
1 & 0 \\ 
0 & -1
\end{array}
\right)  \label{pauli spin matrices}
\end{equation}
These are the well-known Pauli spin matrices; note that the square of each
is the unity matrix.

With this in mind, we proceed to remind ourselves of some facts about the
angular momentum of a 2-particle composite system. ${\bf S_{1}}$ and ${\bf %
S_{2}}$ are the spin angular momenta of the two particles and ${\bf S}{\bf %
\equiv }{\bf S_{1}}$ $+{\bf S_{2}}$. If $s_{1}$ and $s_{2}$ are the spin
quantum numbers for the particles, the total spin quantum number $s$ for the
whole system ranges in integral steps from $\left| s_{1}-s_{2}\right| $ to $%
s_{1}+s_{2}$. For any one $s$, the eigenvalues of $S_{z}$ (the $z$-component
of total spin) range in integral steps from $-s$ to $s$. Finally note that $%
\mid S_{z}=m>$ can be expressed in terms of $\mid S_{1z}=m\prime >$ and $%
\mid S_{2z}=m\prime \prime >$ by referring to the Clebsch-Gordon/Wigner
coefficients.

Now, for the singlet state of two spin-$\frac{1}{2}$ particles, $s=0$ and $%
m=0;$ so the resulting state vector is:

\begin{equation}
\mid \Psi _{singlet}>\,=\,\mid \alpha (1)>\mid \beta (2)>-\mid \beta
(1)>\mid \alpha (2)>  \label{singlet}
\end{equation}

We observe that it is rotationally invariant.

For later use, we will now calculate the correlation function which is
defined as:\footnote{%
Here $\otimes $ denotes the tensor product of operators (over a tensor
product of Hilbert spaces). Roughly speaking, if our composite system is
made up of states of $\psi $ (a system with Hilbert space $\aleph $) and
states of $\psi ^{\prime \text{ }}$(a system with Hilbert space $\aleph
^{\prime }$), then the Hilbert space of the composite system will be the
tensor product $\aleph \otimes \aleph ^{\prime }$.} 
\begin{equation}
C({\bf a},{\bf b})=<\Psi _{singlet}\mid ({\bf \sigma _{1}}\cdot {\bf a}%
)\otimes ({\bf \sigma _{2}}\cdot {\bf b})\mid \Psi _{singlet}>
\label{definition of correl}
\end{equation}

Each dot product of spin with a unit vector gives the component of the spin
in that direction. The correlation function is just the average value of the
product of the spin component for particle 1 with the spin of the component
for particle 2; this can be easily checked by inserting completeness
relations where appropriate.

Let the $z$-axis be along unit vector ${\bf a}$; choose the $x$-axis so that 
${\bf b}$ lies in the $x-z$ plane, and let ${\bf b}$ make an $\theta _{ab}$
with ${\bf a}$. We can then write

\begin{equation}
C({\bf a},{\bf b})=<\Psi _{singlet}\mid (\sigma _{z1}\otimes (\sigma
_{z2}\cos \theta _{ab}+\sigma _{x2}\sin \theta _{ab}))\mid \Psi _{singlet}>
\label{correlation of singlet I}
\end{equation}

From equation \ref{singlet} and \ref{definition of correl}, we can work out
the action of the Pauli spin operators to give:

\begin{equation}
C({\bf a},{\bf b})=-\cos \theta _{ab}  \label{correlation of singlet II}
\end{equation}

We will also need the probability that the spin component measured along the
z-axis is opposite in sign to the spin component measured along the vector
at angle $\theta _{ab}.$ To get this, we first solve the eigenvalue
equation: 
\begin{equation}
S_{\theta _{ab}}\mid vector>=a\mid vector>
\end{equation}
where $S_{\theta _{ab}}=(\sigma _{z2}\cos \theta _{ab}+\sigma _{x2}\sin
\theta _{ab}),$ which gives eigenvalues of \ +1 and -1 with eigenvectors,
respectively: 
\begin{equation}
\binom{\cos [\theta _{ab}/2]}{\sin [\theta _{ab}/2]}\,\text{and }\binom{%
-\sin [\theta _{ab}/2]}{\cos [\theta _{ab}/2]}\text{ }
\end{equation}

Next, decompose the spin up eigenvector, $\binom{1}{0}$ into the above
eigenvectors of $S_{\theta _{ab}}$ and similarly for the spin-down one $%
\binom{0}{1}.$ This yields straightforwardly the probability that the
particle in the pair encountering the z-axis detector will be measured with
opposite eigenvalue to that encountering the detector set to measure the
spin component at angle $\theta _{ab}$; that is, one is looking for \ the
probability that the eigenvalues of the pair are $(+1,-1)$ or $(-1,+1)$).
The probability is:

\begin{equation}
\cos ^{2}\frac{\theta _{ab}}{2}  \label{probability of anti-alignment}
\end{equation}

\subsection{Bell's Theorem: Heuristically}

First, to make the proof accessible to those new to the subject and to
render the problem sufficiently clear to others\footnote{%
Those who have extensive background may want to skip to the mathematical
proof in the next subsection.}, I'll lay out the experiment and then
describe Bell's conclusions from it.\footnote{%
The main line of argument followed here is taken from my layman's level book
``The Science\ Before Science.''\cite{RizziSciBeforeSci} The particulars
have been modified to address physicists more directly.} I use the
incarnation of the EPR paradox described by Bohm and Aharonov\cite
{Aharonov-Bohm57}. Let's consider a pair of electrons in the singlet state.
A set of these quantum mechanically entangled electrons are such that the
mathematical relationships described above obtain between the measured
values of the spin of each electron. The electron spins are precisely
anti-correlated when measured at the same angle.

For concreteness, say the two correlated electrons are emitted from a
nucleus of a particular atom, and they move in opposite directions. The
experimental setup is shown schematically in figure 1. On the left and
right, symmetrically placed around the atoms (nuclei) that emit the
correlated electrons are two detectors. The left (right) detector can be set
at some angle L (R) with respect to some vertical axis defined as ``up,''
i.e. the vertical will be considered zero degrees; this is a mere convention
that does not effect the generality of the results. When set at a given
angle, an electron striking a detector yields a binary value, which we
designate as ``+'' for spin up and ``-'' for spin down, depending on whether
the electron goes, respectively, along the direction in which the angle ray
points or away from it (see figure 1).\footnote{%
Such an apparatus uses a magnetic field and is called a Stern-Gerlach
experiment.} One thus can speak about a binary measured value for the spin
component at any one angle. Quantum mechanics predicts that after a series
of nuclear emission events, each of the resulting pairs of spin measurements
has a probability of $\cos ^{2}[(L-R)/2]$ of being anti-aligned (i.e. one
electron $+$ and its pair $-$), where, note, $L-R$ is the difference between
the two angles (cf. equation \ref{probability of anti-alignment}). For
example, if $L-R=60$ degrees, then there is $75\%$ probability of the two
measurement results being anti-aligned. To see the problem take the
following scenario:

1) With $L=0$ degrees and $R=60$ degrees, take 10 data runs, i.e., wait for
10 nuclear emission events. This experiment results in 10 electrons
impinging on each detector and, hence, 10 outputs from each detector
indicating whether an electron was measured to be up ``$+$'' or down ``$-$''.

2) Now say that instead of the above, we did the experiment by taking 10
data runs $L=0$ degrees and $R=120$ degrees.

3) Fill the data into the corresponding parts of the table below. We, of
course, pick data in such a way as to be in agreement with the predictions
of quantum mechanics.

4) Knowing that the conservation of angular momentum will fix the remainder
of $L=R=0$ degrees, $L=R=60$ degrees and $L=R=120$ degrees portion of the
table (that is when the angles, $L$ and $R$, are the same). In other words,
``$+$'' on one detector implies ``$-$'' on the other and vice-versa. Since
these entries in the table were not obtained by experiment, we print them in
a smaller font.

\[
\begin{tabular}{||c||c|c|c|c|c|c|c|c|c|c||c|c|c|c|c|c|c|c|c|c||c||}
\hline
\multicolumn{11}{|c||}{\bf LEFT DETECTOR} & \multicolumn{11}{|c|}{\bf RIGHT
DETECTOR} \\ \hline
{\bf L} & 1 & 2 & 3 & 4 & 5 & 6 & 7 & 8 & 9 & 10 & 1 & 2 & 3 & 4 & 5 & 6 & 7
& 8 & 9 & 10 & {\bf R} \\ \hline
0 & + & - & + & - & + & - & + & - & + & - & {\tiny -} & {\tiny +} & {\tiny -}
& {\tiny +} & {\tiny -} & {\tiny +} & {\tiny -} & {\tiny +} & {\tiny -} & 
{\tiny +} & {\tiny 0} \\ \hline
{\tiny 60} & {\tiny -} & {\tiny +} & {\tiny -} & {\tiny -} & {\tiny +} & 
{\tiny -} & {\tiny +} & {\tiny -} & {\tiny +} & {\tiny -} & + & - & + & + & -
& + & - & + & - & + & 60 \\ \hline
{\tiny 120} & {\tiny -} & {\tiny +} & {\tiny -} & {\tiny -} & {\tiny +} & 
{\tiny -} & {\tiny -} & {\tiny +} & {\tiny -} & {\tiny +} & + & - & + & + & -
& + & + & - & + & - & 120 \\ \hline
\end{tabular}
\]

Now, if we look at many of the various portions of the table, we see that
they are correlated, as they should be, approximately according to $%
cos^{2}[(L-R)/2]$. Specifically, for $L=0$ degrees and $R=60$ degrees the
electrons in a pair are anti-aligned about $70\%$ of the time, which is as
close as one can get to the predicted $75\%$ with just 10 runs. Similarly,
for $L=0$ degrees, $R=120$ degrees, there is $30\%$ anti-alignment, compared
to the predicted $25\%$.

So, what's the dilemma? Note that the quantum mechanical prediction of the
correlation (which, as defined in this section is $cos^{2}[(L-R)/2]$) does
not depend at all on the absolute angle of either detector; it just depends
on the difference. On the other hand, the table shows that comparing $L=60$
degrees and $R=120$ degrees gives us only $60\%$ anti-alignment where we
needed $75\%$ (i.e. should be $70$ or $80\%$) by the prediction $%
cos^{2}(60/2)$.

Now one may think the answer is simply that all statistics implies some
variation, and with this small a sample, one can expect a lot of variation
from the predicted value. However, try to make the situation better by
manipulating the ``$+$'' and ``$-$'' signs in the table. You cannot; it is
imposed on you by the fact that each angle must give on the average 
$\frac12$%
$+$'s and 
$\frac12$%
$-$'s and the correlation between measurements of all the various angle
combinations must be given by the $cos^{2}[(L-R)/2]$ rule. One cannot
satisfy the correlation for all angles simultaneously. So, how can the
electrons, which are in principle separated by great distance when they're
measured, ``know'' what angle to correlate to? For example, if the two
detectors are set approximately equidistant from the source, then one
measurement will be happening about when the other is; hence, one can set
the time available for one measurement to affect the other arbitrarily
small. Indeed, the experiments already actually done did not leave enough
time for a signal to propagate at the speed of light from one electron to
its pair. Hence, the electrons cannot be initially set to do what they do,
because we've seen that the correlation would not then be what it is
observed to be. This is the idea of Bell's theorem.

\subsection{\protect\bigskip Bell's Theorem: Mathematical Proof}

One nice proof of Bell's theorem, which follows closely Bell's own proof\cite
{BellEPR64}, is given in the appendix (pg 233) of a text by Rae.\cite{RaeQM}
We will follow his proof in a notation slightly modified to be consonant
with the heuristic treatment above. In his and Bell's proof, one assumes
that the result of each measurement depends only on a mathematical
description that includes only what's happening {\it near the electron being
measured} not the far away electron. Assuming this, one then deduces that
the resulting correlations will not agree with experiment.

Specifically, one starts by assuming that the outcome of each measurement at
any angle is only a function of a random variable, '' $\lambda $''
associated with each pair. Bell points out that $\lambda $ can actually be
many variables or function of many variables. Since the measured behavior of
any given electron in the experiment is quantified by $\lambda $ which is
``carried with'' the given electron, assigning $\lambda $ enforces the
locality assumption. In other words, by assigning $\lambda $, we assume the
measurement of one electron doesn't influence the state of its pair, only
the locally determined $\lambda $ does.

\bigskip The proof proceeds as follows using the conventions defined above. $%
\lambda $ is a hidden variable that determines $S_{zleft}$, the $z$%
-component of the spin of the particle moving to the left. More generally,
the first subscript specifies the angle of the spin component measured and
the second subscript specifies which detector the particle is approaching,
the left or the right.

We define $p(\lambda )$ as the probability density so that $p(\lambda
)d\lambda $ is the probability of a pair being produced with a value of $%
\lambda $ between $\lambda $ and $\lambda +d\lambda $. All the probabilities
should add up to one so we lay down a normalization condition: 
\begin{equation}
\int p(\lambda )d\lambda =1
\end{equation}

We now consider an experiment in which the spin components corresponding to $%
S_{zleft}$ and $S_{\phi right}$ are measured on a large number $N$ of
particle pairs. Accordingly, the average of the values of the products $%
S_{zleft}(\lambda )S_{\phi right}(\lambda $) will be given by $C(\phi )$
where 
\begin{equation}
C(\phi )=\int S_{zleft}(\lambda )S_{\phi right}(\lambda )p(\lambda
)\,d\lambda
\end{equation}
This expression is clearly true since it is just the limit of the sum of the
products multiplied by the probabilities.

If we now consider that {\it instead} of doing the previous experiment on
the $N$ electron pairs, we did a slightly different one. We keep the left
apparatus as before, but the right apparatus is instead oriented to measure
the component at an angle $\theta $ to the $z-$axis. We would thus obtain a
similar expression for the quantity $C(\theta )$ and hence

\begin{equation}
C(\phi )-C(\theta )=\int \left[ S_{zleft}(\lambda )S_{\phi right}(\lambda
)-S_{zleft}(\lambda )S_{\theta right}(\lambda )\right] p(\lambda )\,d\lambda
\label{difference in correlations a}
\end{equation}

Since the sum of the same (measured) component of spin of the left particle
and the right particle is always zero, we get 
\begin{eqnarray}
S_{\theta right}(\lambda ) &=&-S_{\theta left}(\lambda )
\label{anti-correlation at angle theta} \\
S_{\phi right}(\lambda ) &=&-S_{\phi left}(\lambda )
\label{anti-correlation at angle phi}
\end{eqnarray}

Thus:

\begin{eqnarray}
C(\phi )-C(\theta ) &=&-\int S_{zleft}(\lambda )\left[ S_{\phi left}(\lambda
)-S_{\theta left}(\lambda )\right] p(\lambda )\,d\lambda
\label{difference in correlations b} \\
&=&-\int S_{zleft}(\lambda )S_{\phi left}(\lambda )\left[ 1-S_{\phi
left}(\lambda )S_{\theta left}(\lambda )\right] p(\lambda )\,d\lambda
\label{difference in correlations c}
\end{eqnarray}

The last step follows because $S_{\phi left}(\lambda )=\pm 1$

Taking the absolute value of both sides we get:

\begin{eqnarray}
\left| C(\phi )-C(\theta )\right| &=&\left| \int S_{zleft}(\lambda )S_{\phi
left}(\lambda )\left[ 1-S_{\phi left}(\lambda )S_{\theta left}(\lambda )%
\right] p(\lambda )\,d\lambda \right|  \label{absolute of correlation diff a}
\\
&\leqslant &\int \left| S_{zleft}(\lambda )S_{\phi left}(\lambda )\left[
1-S_{\phi left}(\lambda )S_{\theta left}(\lambda )\right] p(\lambda )\right|
\,d\lambda  \label{absolute of correlation diff b}
\end{eqnarray}

Since $\left| S_{zleft}(\lambda )S_{\phi left}(\lambda )\right| =1$ ,
equation \ref{absolute of correlation diff b} can be written as

\begin{eqnarray}
\left| C(\phi )-C(\theta )\right| &\leqslant &\int \left[ 1-S_{\phi
left}(\lambda )S_{\theta left}(\lambda )\right] p(\lambda )\,d\lambda
\label{absolute of correlation diff c} \\
&\leqslant &1+\int S_{\phi left}(\lambda )S_{\theta right}(\lambda
)p(\lambda )\,d\lambda  \label{absolute of correlation diff d}
\end{eqnarray}

The integral in equation \ref{absolute of correlation diff d} is simply the
correlation function between the measured value of two spin components that
are at an angle of $(\theta -\phi )$ with each other and is therefore equal
to $C(\theta -\phi )$. Thus we can write equation \ref{absolute of
correlation diff d} as: 
\begin{equation}
\left| C(\phi )-C(\theta )\right| -C(\theta -\phi )\leqslant 1
\label{bell's inequality}
\end{equation}

This is one of the forms of the so-called Bell's inequalities. Substituting
the quantum mechanically correct correlation function for $C(\alpha )=-\cos
(\alpha )$ from equation \ref{correlation of singlet II}, one finds a whole
range of angles at which the inequality fails. For example, take the case we
used in the section above where: $\theta =120^{\circ },\phi =60^{\circ }.$
In this case, Bell's inequality is written:

\[
\frac{3}{2}\leqslant 1 
\]
This statement is clearly false. Hence, it is thus proved that the
predictions of quantum mechanics, which are the ones found in experiment,
cannot be true if the other assumptions we've made are true. By a reductio
ad absurdum, we concluded that what we take to be our key assumption must be
wrong, and each electron must be dependent on the state of the disconnected
other.

Many in fact conclude that nature is non-local from this. They say there
must be action at a distance, meaning instantaneous action of one body on
another. Others, trying to avoid conceding action at a distance, conclude
that spin properties and other properties that have non-commuting variables
such as position and momentum, do not exist until they are observed; still
many say there is no such escape.

\section{The Meaning of Bell's Theorem\label%
{Section: The Meaning of Bell's Thereom}}

\bigskip Now, locality is a cherished principle still even today in most of
physics; most would rather not give it up. To certify that there is no
escape route in such a proof, one must make sure it contains no hidden
assumptions. We will do this check, but first there are a couple signs, not
proofs, that we might be on the wrong track in concluding that quantum
mechanics implies non-locality. First, it is well known that quantum
mechanics does not imply action at a distance in one very pragmatic sense.
Philippe Eberhard showed that within quantum mechanics, there cannot be
communication of information faster than the speed of light.\cite{Eberhard78}
In other words quantum mechanics does not violate the letter of the law of
special relativity. The second sign comes from the statistical nature of
quantum mechanics.

\subsection{Signs of Locality}

Quantum mechanics is a statistical theory. The use of statistics is an
admission of lack of knowledge. We only use statistics when we do not know
(by choice or by reason of some impassable obstacle) the details of a
particular phenomenon under consideration. In so doing, we leave out
specifics and settle for averages. For example, in a coin toss, we say that
there is a probability of one in two that it will be heads. In saying this
we are relating very little about the individual coin toss. We are relating
something about a large group of coin tosses. This allows us to not be
concerned with the details of each toss, which are very complicated. Hence,
it would be a great error to conclude that because we have an element of
randomness in our description that the flipping of our coin is intrinsically
random. This is to forget that we have deliberately left these details out.

Further, in taking averages as one does in statistics, it is easy to come to
conclusions that take root in the failure to remember what you've included
and what you've not included in your theory. Consider the case of a man who
drowned in a lake with an average depth of two inches. Each member of a
group of non-swimmers that walked across this lake could be told that his
chance of drowning is very small. For anyone that actually does drown such a
statement will have no importance. Specifics, not generalities, are in the
end what reality is about. When we describe a group of things under
consideration by statistics we are admitting our deficit of knowledge about
those things, and yet making use of that knowledge to say what little we can
about them.

Since quantum mechanics makes use of statistics, we need to be wary that the
individuals may be left out of the account. Hence, concluding that there is
action at a distance, which means one electron of a given pair is acting
instantaneously on the other of the same pair, from quantum mechanics seems
problematic.

In particular, since the Eberhard proof says the statistics of measurements
cannot be altered at a distance, i.e. they behave locally, and since it is
measurement statistics that are directly described by quantum mechanics, it
would be odd if the individual behavior were non-local. This oddity has not
gone unnoticed, but it is nonetheless a key sign that something might be
amiss with the non-locality proofs.

To resolve these issues, we now look more closely at the experiment and the
proof.

\subsection{A Second look at the Experiment}

We will soon see that in our forming of the dilemma, we implicitly assumed
that we could measure the spin state of one electron at two different angles
without one measurement affecting the outcome of the other. The uncertainty
principle testifies that we are allowed no such ability. Indeed, the whole
standard interpretation of Bell's theorem hinges on the possibility of
making measurements involving at least three different angles, which, in
turn, means one must consider the measurements at two angles for one
electron simultaneously. How so?

Recall that the measurements above were taken by changing the angle $R$ of
detection of the right detector from $60$ degrees to $120$ degrees, but the
measurement at each angle was taken as if the experimenter had not done the
other (see step one and two above). We did the first experiment with $R=60$
degrees and then we said what happens if we did not do it, but did do $R=120$
degrees. Indeed, we can say that one of these two experiments is really
simply a hypothetical experiment, because we are in some sense refusing to
allow {\em both} actually occurred. Yet, in our analysis we considered the
data as if the experiments were both really done; thus, we implicitly
assumed in our reasoning that it would yield the same result if both were
really done on the same electron. This is a leap of logic; it leaps over the
implicit assumption that one measurement would not affect the other.

In short, as long as we respect our {\it inability} to measure two {\it %
different} spin components of the {\it same} electron (not a statistical
electron) {\it without} one measurement interfering with the result of the
other, the conclusion will not follow; that is, we will not be forced to
assume action at a distance.

\subsection{Reexamining the Mathematical Proof}

In the mathematical proof given above, the case is clear. We twice assign 
{\it measured} values at two different angles to one electron. To manifest
the hidden assumptions, take the $\lambda $ as an index or label marking the
pair under consideration.

Note that each electron measurement is associated with one value of the
hidden variable. However, it may be that the association is not one-to-one,
so that there is more than one electron with the same $\lambda .$ Indeed,
one could assume that the measured spin is a bounded smooth function of the
hidden variable and thus will be repeated after enough runs, so that one can
avoid having to measure the same one twice. However, such an assumption is
by no means forced on you. Each pair can be completely unique; indeed each
is already unique at least to the extent that each pair succeeds another in
coming out of the source, and they come out of different nuclei in our
scenario.

In any case, we are free to take the $\lambda $ as an index marking which
pair is under consideration. By assigning, in equation \ref{anti-correlation
at angle theta} , $S_{\theta right}(\lambda )=-S_{\theta left}(\lambda ),$
one is implicitly asking for the measured spin component for particle ``$%
left"$ at both $\theta $ and $0\unit{%
{{}^\circ}%
}$ (that is along the $z-axis$). $\ $This can be seen clearly in equation 
\ref{difference in correlations b}, where one sees that both $%
S_{zleft}(\lambda )$ and $S_{\phi left}(\lambda )\ $are used. The parallel
equation \ref{anti-correlation at angle phi}, $S_{\phi right}(\lambda
)=-S_{\phi left}(\lambda ),$ similarly implies one is measuring the spin
component for a single particle for both $\phi $ and $z.$ Indeed, the
problem already appears in equation \ref{difference in correlations a} where
both $S_{\phi right}(\lambda )$ and $S_{\theta right}(\lambda )$ are needed.
Two hypothetical experiments are involved in getting these measured values.
Either they are actually done or they are not actually done. If they are not
done then we do not have the results. If they are done then first interferes
with the result of the second; the second is no longer governed by $%
S(\lambda )$ in the same simple way.

Again, the only way we could know both values of the components of the spin
is if the measured values were to be independent of each other, i.e. one not
affected by the other. This is a fine starting assumption as long as one
recognizes it as such. Then one is trying to prove that one cannot
simultaneously know to arbitrary accuracy, by measurement, the value of two
different spin components of one electron. If this is the case the proof
goes through flawlessly. That is, one can end up either concluding that such
simultaneous exact knowledge by measurement cannot occur or that there is
action at a distance; we choose the former, which squares with Heisenberg's
uncertainty principle for non-commuting variables.

Of course, one's inability to simultaneously measure two aspects of a thing 
{\it exactly} does not mean that those aspects cannot exist simultaneously,
so there is no real contradiction in this case. Indeed, if one chooses to
make the possibility of exact measurement a necessary condition for the
existence of a thing then {\it our} {\it mere} abilities determine nature.
That is, we don't even have to exercise our abilities, {\it we} somehow {\it %
passively, by our mere existence,} cause things to be. Einstein insightfully
remarked that all science depends on the negation of such a proposition,
pointing to science's implicit reliance on the fact that the world exists
independent of us.

So, how do we interpret the EPR experiment described above. First, we say
the measurement of one of the electrons gives its ``measure state''
described by quantum mechanics. Once we know that state for a given angle,
we can use our statistical knowledge of that ''measure state'' to predict
the outcomes of further measurements on that electron. Further, by using the
quantum predictions of the correlation of the pair, we also instantly know
that the other electron will be in the same ``measure state'' with respect
to the given angle; this fixes probabilities for measurement at other
angles. The electron pairs carry their correlation from their emission from
a common nucleus, but we are ignorant as to the specifics of that process
and how it relates to the results of our measurements because of the
limitations of our physical knowledge, at least at present.

\section{Conclusion}

\qquad Its been said that EPR is the most cited paper in the literature.\cite
{LaloeReviewArticle} This is an odd fact in one way, because it is often
considered that such topics are on the boundary between physics and
philosophy. In any case, EPR and Bell's theorem touch the heart of what
physics is about. It reveals its strength and limitations. The key strength
of modern physical theories, their ability to accurately predict
experimental outcomes even when our understanding of the theory and
intuition about it are underdeveloped and confused or even somewhat
contradictory, is manifested by the quantum mechanics of Bell's theorem
discussed above. In this case, its strength also appears to be its weakness,
because the theory's accuracy has induced serious and otherwise reasonable
people to doubt the reality of the world independent of us by taking serious
various interpretations of the theory.

The discussion above reveals somewhat the genesis of these kind of
confusions. Confusions can arise when one does not give adequate attention
to the step of advancing from the already existing mathematical structure of
a theory and the accompanying physical situations that it describes{\it \ to}
new theories and understandings. In particular, this step requires careful
thinking about which direction one would like to go. There are two main
paths to consider. If a physicist wants to delve more deeply into a theory's
mathematical structure to find a more powerful theory that incorporates more
experimental data and circumvents current weaknesses, he should continue in
the generic line of thinking that brought the equations to the fore. This
emphatically does {\em not} mean he should follow the exact same
mathematical method and insights and avoid new ground; such is the
definition of a sterile approach. I mean that philosophical sorts of
considerations should not be allowed to control his thinking too much. By
contrast, when the questions are about the ontological realities of the
entities he is describing, he should step back a bit further and look
carefully at the larger assumptions.

Not doing so can lead to unlikely or absurd conclusions. We saw this happen
in our discussion of Bell's theorem; generically, we saw what looked fine
mathematically was actually impossible physically.

It is instructive to recap the key point of our Bell's theorem discussion.
To facilitate the discussion re-write $S_{\phi left}(\lambda ),$ the
measured value of the spin component in the $\phi $ direction of a
particular particle as it moves toward the left detector, as: $S_{left}(\phi
,\lambda )$. This way of writing it fully manifests that $S$ is a function
of two quantities. For a given electron of given $\lambda ,$ this function
can be a smooth function of $\phi .$ That is, mathematically, one can think
of all values of the function $S$ existing at once, as does, for example, a
plot of the function as a two dimensional surface in three dimensions.
Despite this mathematical fact, however, physically, for fixed $\lambda $,
one can only know the function's value for one value of $\phi $ at a time.
Why? Because the act of measuring the electron in some way alters the other
values, putting the particle in a different ``measure state'' therefore
requiring a different functional description say $S_{left}(\phi _{1st},\phi
_{2nd},\lambda )$, which now depends on what angle was measured first. Like
a match, it can only be lit once, but can potentially lit many different
ways: e.g., striking on the side of the box, on a table surface, in the
middle by a second match etc...

\qquad In the particular case considered above, we can determine two angles
for one electron by measuring it and its pair each once. For example, I can
measure the left particle at $\phi =0^{\circ }$ and the right particle for $%
\phi =60^{\circ }$ and will thus know the ``measure state'' of each. But now
neither is in the same ``measure state'' as before the measurement, so the
left will, in general, {\em not} measure at $60$ degrees what the right
measured and vice versa at $0$ degrees.

I use the term ``measure state'' above to indicate that the reduction in the
wave packet is simply an increase of our knowledge of the system. It is
suggestive of the following interpretation of quantum measurement. Because
the statistical nature of quantum mechanics, which is in turn due to our
ignorance of the system, we must have one piece of information about the
system before we can say anything further than just the correlation, i.e.
the probability of the left detector measuring spin up at a given angle,
given a measurement at a spin component at a given angle on the right
detector. The jump is a jump in our knowledge, not in the reality of the
thing measured. To manifest the absurdity of that position, the situation
might be compared to the classical situation in which one needs the initial
velocity and position of a projectile, say a ball, before one can use the
laws of Newton to predict where it will land. One might say that until one
observes the ball at a particular time, the ball is everywhere. After all,
one needs measurements to say anything about a particular real ball.
Equations giving statistical outcomes need initial data as well, just of a
different type and interpretation. In either case, one can conclude
measurement causes reality instead of revealing something of it. Of course,
such a statement gets the facts exactly backwards; we know, for example,
that the ball exists first and then experiment to obtain the laws of
Newtonian physics that describe the dynamics of the ball.

The insights and arguments laid down here I first produced and wrote
independently in the previously cited manuscript ``the Science\ Before
Science,'' and only later discovered that T. Brody had already said some of
these things, though without all the detailed explications outlined above.
Brody discusses the key issues\cite{BrodyBook}\cite{Brody79} and even
concludes with an example of an equation that reproduces the Bell's
inequality.\footnote{%
He also references other such examples; confer \cite{Scalera83}\cite
{Scalera84} and \cite{Notarrigo84}.} He calls the inability of
simultaneously knowing two variables that often attends violations of Bell's
inequality the joint measurability condition.

Some have either tried to avoid the simultaneous measurability condition or
seem to have avoided it, but end up implicitly assuming it any way. Certain
popular treatments sometimes seem to avoid the condition, but also
implicitly assume it.

\subsection{Acknowledgments}

\qquad I'd like to thank Nicholas Teh of Princeton University for
contributing most of section 2.1 (the qm of the electron pair singlet
state), and assisting with\ exposition of the history in the introduction. I
also thank him for his comments on the style and presentation of the
arguments in the manuscript. Thanks to Stephen Barr and Juhan Frank for
their comments on the paper and special thanks to Michael Romalis for
helpful discussions on the manuscript.

\bibliographystyle{unsrt}
\bibliography{btxdoc,thsisprd}
\pagebreak

\begin{figure}
\caption{The experimental setup as seen from an off-perpendicular angle. 
A particle decays into a pair of spin-entangled electrons, one
moving to the right, the other to the left; each moves towards its own detector. 
One sets each detector to determine the spin component at a given an angle;
spin along the angle vector is written as ``+" and spin opposite as ``-"}
\label{EPR experiment}
\end{figure}
%

\end{document}